\documentstyle[aps,psfig]{revtex}
\begin{document}
\draft
\title{
\hfill{\small {\bf MKPH-T-00-13}}\\
{\bf 
Three-body analysis of incoherent $\eta$-photoproduction
on the deuteron in the near threshold region}
\footnote{Supported by the Deutsche Forschungsgemeinschaft (SFB 443)}}
\author{A. Fix$^a$ and H. Arenh\"ovel$^b$}
\address{$^a$Tomsk Polytechnic University, 634034 Tomsk, Russia}
\address{$^b$Institut f\"ur Kernphysik,
Johannes Gutenberg-Universit\"at Mainz, D-55099 Mainz, Germany}
\date{\today}
\maketitle

\begin{abstract}
A three-body calculation of the reaction $\gamma d\to\eta np$ in the
energy region from threshold up to 30 MeV above has been performed.
The primary goal of this study is to assess the importance
of the three-body aspects in the hadronic sector of this reaction.
Results are presented for the $\eta$-meson spectrum as well as for
the total cross section. The three-body results differ significantly
from those predicted by a simple rescattering model in which only first-order
$\eta N$- and $NN$-interactions in the final state are considered.
The major features of the experimental data are well reproduced although right
at threshold the rather large total cross section could not be explained.
\end{abstract}

\section{Introduction}
In this letter we would like to present theoretical results
for total and differential cross sections of the reaction
$\gamma d\to \eta np$ close to the threshold region.
The need for an improved theory for describing the low-energy
$\eta$-production on the lightest nuclei stems from the recent precise
measurements of such processes on light nuclei using
hadronic and electromagnetic probes. The experimental results are
substantially larger than a mere phase-space calculation near
threshold \cite{Calen,Metag}.
It is reasonable to associate this effect with a rather strong
attractive force which governs the dynamical features of the $\eta
NN$-system in the low-energy regime. In a previous paper \cite{FiAr1}
we found that the $\eta NN$-dynamics allows the existence of virtual s-wave
three-body states near zero energy. The natural extension
of these findings would be to investigate to what degree such
states may influence the processes considering the $\eta NN$-system in
different spin-isospin channels.
Of particular interest are the reactions of coherent ($\gamma d\to \eta d$)
and incoherent ($\gamma d\to \eta np$) $\eta$-photoproduction on the deuteron,
which recently have become the subject of attention of different
experimental collaborations \cite{Metag,Hoff,Krusche2}.

As for the coherent process, the investigation of this channel
within a three-body approach has been performed in \cite{Ueda}.
Unfortunately, due to the smallness of the coherent $\eta$-photoproduction
cross section resulting in extremely small $\eta$-meson yields, only
qualitative statements about the size of the experimental cross section of
the coherent reaction is available at present.

In this work, we focus our attention exclusively on the incoherent
$\eta$-photoproduction where the deuteron breaks up. Recent precision
measurements of the break-up channel~\cite{Metag} require a more sophisticated
theoretical analysis. Indeed, this reaction has already been considered
in a previous paper~\cite{FiAr} within the so-called rescattering model
as a first step beyond the simple impulse approximation (IA). In this model,
final state interaction is taken into account by including the single
rescatterings between two of the final particles as most important
correction to the IA. As was shown in \cite{FiAr}, without such rescattering
effects a strong suppression of $\gamma d\to\eta np$ cross section was found
because the IA requires a large spectator nucleon momenta for producing a
low-energy $\eta$-meson on the deuteron.
For this reason, the inclusion of first order $\eta N$- and $NN$-interactions
provides a mechanism to distribute the high momentum between two nucleons
and thus avoids this suppression. In view of the importance of the first-order
rescattering it is natural to ask about the role of higher order rescattering
terms. Indeed, in a system of three low-energy particles with rather strong
attractive forces, one can expect a considerable overlap of their wave
functions. Therefore, there
is no guarantee that the contributions from higher order $\eta N$- and
$NN$-rescatterings will be small. This fact distinguishes the present reaction
from those with pions where the smallness of the pion-nucleon
scattering length allows one to take into account only the lowest order
$NN$-interaction in the final state just above threshold \cite{Laget}.
Thus, the rescattering approximation can hardly be considered as
a satisfactory method for describing low-energy $\eta$-photoproduction
on light nuclei. It must also be noted that the successive
inclusion of the next higher order terms of the multiple scattering
expansion is not well advised, because the corresponding Neumann series is
expected to converge very slowly due to the proximity of the virtual
three-body poles mentioned above.
As a consequence, the inclusion of terms of quite high order
will be required rendering such an approach inadequate.
These considerations point to the necessity of a three-body approach,
where the final state interaction is included to all
orders.

In the present paper, the reaction $\gamma d\to \eta np$ is considered
as a three-body problem in the hadronic sector.
The formalism is based on a separable s-wave
approximation for the two-body potentials. This approach may be
justified by the fact that in the energy region under consideration
the $\eta N$-resonance $S_{11}$(1535) as well as the $^1S_0$ and $^3S_1$ poles
of the $NN$ scattering matrix determine essentially the driving $\eta N$-
and $NN$-forces. Our primary goal is to investigate the main properties
of the incoherent $\eta$-photoproduction on the deuteron near threshold. We
also compare our results with those predicted by the first order rescattering
approximation in order to assess the shortcomings of this method.

In the Sect.~\ref{sect1}
the three-body equations for the $\gamma d\to \eta
np$ transition amplitude are derived. In Sect.~\ref{sect2}
we calculate the meson spectrum for a
fixed photon energy and compare our results for the total cross section
with available experimental data.

\section{General formalism}\label{sect1}

Here we will briefly describe the main points of the
formal part, and comment on the most important approximations involved in our
calculation. A more detailed and rigorous
development of the formalism as well as the method for
the numerical solution of the three-particle integral equations will be
presented elsewhere. Firstly, we note that the three-body concept
of the $\eta$-photoproduction on a deuteron possesses in principle
a straightforward
solution of the three-body multichannel problem where the coupling
between the channels $\gamma NN$ and $\eta NN$ is taken into account.
However, since we will restrict ourselves to the first order only in the
electromagnetic interaction, it is more appropriate to
obtain the dynamical equations of our model starting from the
Faddeev formalism for the pure hadronic $\eta NN$ system.
Our approach with respect to the hadronic part is essentially based
on the separable approximation scheme of Alt, Grassberger and
Sandhas~\cite{AGS} with an
isobar ansatz for the $\eta N$- and $NN$-interactions. As is
mentioned above, this ansatz should represent a fair approximation of
the real dynamics in the energy region examined here. We use the following
notations for the basic three-body channels:
\begin{equation}
\begin{array}{cl}
N^* &\mbox{ for the }\eta N\mbox{-isobar plus
spectator nucleon}, \cr
d &\mbox{ for the $NN$-isobar plus spectator meson}.\cr
\end{array}
\end{equation}
and analogous notations for the two-body subsystems.
Then the separable input
\begin{equation}\label{ti}
t_i=|f_i\rangle\tau_i\langle f_i|\ \quad (i=N^*,d)
\end{equation}
allows one to obtain two coupled operator equations for the
rearrangement amplitudes
\begin{eqnarray}\label{eqXh}
X_d &=& 2Z_{dN^*}\tau_{N^*}X_{N^*}\,, \\
X_{N^*} &=& Z_{N^*d}+Z_{N^*d}\tau_dX_d+Z_{N^*N^*}\tau_{N^*}X_{N^*}\,.
\nonumber
\end{eqnarray}
The amplitudes $X_d$ and $X_{N^*}$ determine the transitions from
the initial on--energy-shell
state $\eta d$ to the quasi-two-body final states $\eta
d$ and $NN^*$, respectively.
The potentials $Z_{ij}$ have a conventional meaning, namely
\begin{eqnarray}
Z_{ij}=\langle f_i|G_0|f_j\rangle\,,
\end{eqnarray}
with $G_0$ being the free three-body Greens function.

\begin{figure}
\centerline{\psfig{figure=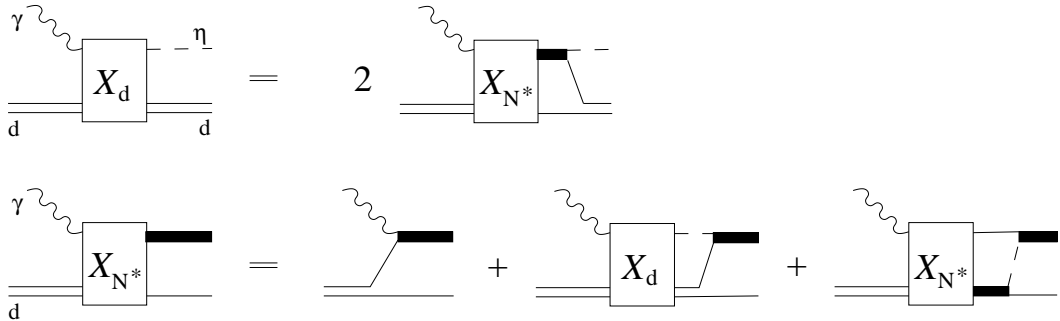,width=14cm,angle=0}}
\vspace{.5cm}
\caption{
Diagrammatic representation of the three-body equations
for $\eta$-photoproduction on the deuteron.
}
\label{fig1}
\end{figure}

Turning now to the photoproduction process $\gamma d\to\eta np$, we treat
the electromagnetic interaction perturbatively, keeping only the terms
up to first order in the e.m.\ coupling. Therefore, in order to introduce
the electromagnetic part, one just has to replace in (\ref{eqXh}) the
$\eta$-meson by the photon in the entrance channel.
This yields then the coupled equations which are formally similar to
(\ref{eqXh})
\begin{mathletters}\label{eqX}
\begin{eqnarray}
X_d &=&
2Z_{dN^*}\tau_{N^*}X_{N^*}\,, \\
X_{N^*} &=& B_{N^*}+
Z_{N^*d}\tau_dX_d+Z_{N^*N^*}\tau_{N^*}X_{N^*}\,.
\end{eqnarray}
\end{mathletters}
Here the ``born term'' $B_{N^*}$ is determined as
\begin{equation}
B_{N^*}=\langle f^{(\gamma)}_{N^*}|G_0^{(\gamma)}|f_d\rangle\,,
\end{equation}
where $G_0^{(\gamma)}$ is the free propagator in the
$\gamma NN$-sector and $|f^{(\gamma)}_{N^*}\rangle$ is the
electromagnetic vertex function for the transition $\gamma N\to N^*$.
Equations (\ref{eqX}) are represented diagrammatically in
Fig.~\ref{fig1}. Clearly, the impulse approximation is equivalent
to the replacements $X_d\to 0$ and $X_{N^*}\to 0$ in the right-hand sides
of (\ref{eqX}), i.e.,
\begin{mathletters}\label{IA}
\begin{eqnarray}
X_d^{\mbox{\scriptsize IA}} &=& 0\,, \\
X_{N^*}^{\mbox{\scriptsize IA}} &=& B_{N^*}\,.
\end{eqnarray}
\end{mathletters}
In order to obtain the first-order rescattering approximation, in which
only the single $\eta N$- and $NN$-scattering terms are retained, one has
just to set $X_d= 0$ and $X_{N^*}= B_{N^*}$. Then one has
\begin{mathletters}\label{Rsct}
\begin{eqnarray}
X_d^{\mbox{\scriptsize resc}} &=&
2Z_{dN^*}\tau_{N^*}B_{N^*}\,, \\
X_{N^*}^{\mbox{\scriptsize resc}} &=&
B_{N^*}+Z_{N^*N^*}\tau_{N^*}B_{N^*}\,.
\end{eqnarray}
\end{mathletters}

After projecting (\ref{eqX}) and correspondingly (\ref{IA}) and
(\ref{Rsct}) onto the $L=0$ states, we end up with a system of
one-dimensional integral equations in momentum space, which can be
solved by matrix inversion. The well known problem of logarithmic
singularities by the rearrangement terms $Z_{N^*N^*}$ and
$Z_{N^*d}$ was avoided using the contour deformation method of~\cite{AAY}.

The transition matrix element for the break-up process
$\gamma d\to\eta np$ may be expressed in terms of
the rearrangement amplitudes (\ref{eqX}) as follows
\begin{eqnarray}\label{eqX0}
\langle\vec{p}_1,\vec{p}_2,\vec{q}\,|T(W)|\vec{k}\,\rangle
&=&
\langle\vec{p}_{\eta 1}|f_{N^*}\rangle\
\tau_{N^*}(W-E_{p_2}-\frac{p_2^2}{2M_{N^*}})\
\langle\vec{p}_2|X_{N^*}(W)|\vec{k}\,\rangle\,
\nonumber \\
&+&
\langle\vec{p}_{\eta 2}|f_{N^*}\rangle\
\tau_{N^*}(W-E_{p_1}-\frac{p_1^2}{2M_{N^*}})\
\langle\vec{p}_1|X_{N^*}(W)|\vec{k}\,\rangle\,
\\
&+&
\langle\vec{p}_{12}|f_d\rangle\
\tau_d(W-\omega_\eta-\frac{q^2}{4M_N})\
\langle\vec{q}\,|X_d(W)|\vec{k}\,\rangle\,.
\nonumber
\end{eqnarray}
Here, $W$ denotes the total three-body c.m.\ energy and $\vec{k}$
the photon momentum. Furthermore, the three-momenta of the
final nucleons and $\eta$-meson are denoted by $\vec{p}_1$, $\vec{p}_2$
and $\vec{q}$, and their total energies by $E_{p_1}$, $E_{p_2}$
and $\omega_\eta$, respectively.
The propagators $\tau_{N^*}(E_{N^*})$
and $\tau_d(E_d)$ depend on the invariant energies of the
corresponding two-body subsystems. Their explicit form is given in
\cite{FiAr1}. The arguments of the vertex functions
$\langle\vec{p}\,|f_i\rangle$ are the respective two-body relative
momenta, whose nonrelativistic expressions read ($i=1,2$)
\begin{equation}
\vec{p}_{\eta i}=\frac{\vec{q}M_N-\vec{p}_im_\eta}{M_N+m_\eta}\,,\ \quad
\vec{p}_{12}=\frac{\vec{p}_1-\vec{p}_2}{2}\,.
\end{equation}
In the actual calculations, only the s-wave configuration
$(J^\pi;T)$=$(0^-;1)$ of the $\eta NN$-system was taken into
account. As was discussed in \cite{FiAr}, due to the spin-isospin
selection rules, this state gives the major contribution to the
$\gamma d\to\eta np$
cross section. We also use the nonrelativistic kinetic energies for all
three final particles, since we restrict ourselves to the
near-threshold region where this approximation is well
justified. The contribution from pion exchange in the
rearrangement potential $Z_{N^*N^*}$ was ignored in view of its
insignificance as was shown in \cite{FiAr1}. The parametrization of
the hadronic separable $t$-matrices (\ref{ti}) is given in detail in
\cite{FiAr1}. For the s-wave singlet and triplet $NN$-states we use
the fit of Yamaguchi~\cite{Yamag}. The $N^*$ parameters
were chosen in such a way that the main hadronic width of the $S_{11}$(1535)
resonance, $\Gamma_{N^*\to\pi N} \approx \Gamma_{N^*\to\eta N} \approx$\,
75\,MeV\cite{PDG98}, is reproduced. As for the electromagnetic vertex
$|f^{(\gamma)}_{N^*}\rangle$, it is natural to associate it with the
$\gamma N\to S_{11}(1535)$ helicity amplitude $A^{(N)}_{1/2}$ which
is of great experimental importance. Thus one has
\begin{equation}
\langle \vec{p}\,|f^{(\gamma)}_{N^*}\rangle\
=A_{1/2}^{(N)}\sqrt{\frac{M_{N^*}}
{M_{N^*}+M_N}}\,.
\end{equation}
Since we restrict our consideration to the $\eta NN$-state with
isospin $T=1$, only the dominant isovector part of the $\gamma N\to
S_{11}(1535)$ transition has been taken into account. The
corresponding amplitude was derived as follows. First we have determined
the helicity amplitude in the proton channel $A_{1/2}^{(p)}$ by fitting
the experimental cross section data for the elementary
process $\gamma p\to\eta p$ of~\cite{Krusche1}.
Then the isovector part $A_{1/2}^{(v)}$ was extracted as
$A_{1/2}^{(v)}=0.9A_{1/2}^{(p)}$ which is compatible with different
analyses~\cite{Hoff,Krusche2,FiAr}.

Finally, we present here the formula for the exclusive c.m.\ cross
section used in the present calculation
\begin{equation}\label{crsec}
d\sigma^{c.m.}=\frac{1}{(2\pi)^5}\frac{E_d\, \omega_\eta E_{p_1} E_{p_2}}{W}\,
\frac{1}{2}\,|\langle\vec{p}_1,\vec{p}_2,\vec{q}\,|T|\vec{k}\,\rangle|^2
\,d\Omega_\eta \,d\omega_\eta \,d\phi_{\eta 1} \,dE_{p_1}\,,
\end{equation}
where $E_d$ is the deuteron energy and
$\phi_{\eta 1}$ is the azimuthal angle of the nucleon momentum
$\vec{p}_1$ counted from the $\vec k$-$\vec q$ plane 
in the frame with the z-axis directed along $\vec{q}$.
The factor $\frac{1}{2}$ appears after summing and averaging over
the spin projections. The semi-inclusive cross section
$d\sigma^{c.m.}/d\omega_\eta$ and the total cross section $\sigma_{tot}$
considered below are obtained from (\ref{crsec}) by appropriate
integrations.

\section{Results and discussion}\label{sect2}

We would like to beginn our discussion with the $\eta$-meson spectrum
$d\sigma^{c.m.}/d\omega_\eta$ as function of the $\eta$ c.m.\ energy,
shown in Fig.~\ref{fig2} for two different photon energies.
Its specific form was discussed already in \cite{FiAr}. A strong s-wave
attraction between the two nucleons at very low relative energy shifts
visibly a major part of the spectrum towards higher energies, resulting in
a rather pronounced peak near the boundary of the phase space at the
maximum of the available meson energy, where the relative energy of the
two nucleons is close to zero. It is obviously a manifestation
of the large $NN$ singlet scattering length, i.e., of the antibound
$^1S_0$-state. As our calculation shows, the first-order rescattering
approximation underestimates substantially the complete three-body
result just above threshold, but overestimates it at higher energies.
Although the form of the spectrum is not changed qualitatively, the
peak is much more pronounced in the complete calculation.

The difference between the results of the first-order rescattering and
the complete three-body calculation is most apparent in the energy
dependence of the total cross section shown in Fig.~\ref{fig3}, where
we compare it also with the inclusive $\gamma d\to \eta X$ data from
\cite{Metag}. As shown in \cite{FiAr} and noted in the introduction,
rescattering gives a very important contribution in particular just
above threshold, where it leads to a strong enhancement of the meson
yield. However, truncating the multiple scattering expansion after the
first order gives only a qualitative description of this effect.
Inclusion of rescattering to all orders within the three-body approach
results in a considerably stronger effect in the near threshold region
so that the energy dependence of the cross section becomes slightly
concave (solid curve ``1'' in Fig.~\ref{fig3}),
which is not characteristic for photoproduction
reactions with more than two particles in the final state. We associate
this effect with the virtual $\eta NN$-pole found in~\cite{FiAr1} in the
$(J^\pi;T)=(0^-;1)$-state which pulls down the essential part of the meson
yield to the near-threshold region.

When comparing our results with the experimental data we see, first of all,
that the three-body calculation constitutes a very important improvement
of the theory which reproduces fairly well the qualitative features of the
data in the near threshold region up to about 650 MeV photon energies,
in particular with respect to the energy dependence. On the other hand,
one notes right at threshold still a substantial underestimation of the data
by the theory. One reason for this disagreement might be, of course,
a contribution of the coherent channel $\gamma d\to \eta d$ to the
experimental inclusive yield. However, to ascribes the whole difference to the
coherent process alone would mean to assign a rather large cross section
of about 100 nb at $E_\gamma$=635 MeV to the coherent process.
While such a value is not in conflict with the recent measurements
of \cite{Metag} it nevertheless
overestimates substantially the corresponding PWIA-calculations for the
case $A_{1/2}^{(v)}=0.9A_{1/2}^{(p)}$ \cite{Breit}. In order to reach
a more definite conclusion one needs, of course, more precise data for
the total cross section of the coherent reaction. Finally, with respect
to the underestimation of the data at energies above 650 MeV, one should
keep in mind that we have included in the calculation only the $\eta NN$-state 
with total orbital momentum $L$=0. Thus at higher photon energies, where the 
higher partial waves are needed to fill the available phase space, our 
calculation cannot be considered to be realistic.

In order to complete our
discussion, we have investigated the sensitivity of the results to the
strength of the interaction, i.e., to a
variation of the $N^*$-parameters. As is shown in \cite{FiAr1} the
$\eta NN^*$ and $\pi NN^*$ coupling constants influence the position of the
$\eta NN$ $S$-matrix poles rather strongly and thus may also
affect the cross section value. Indeed, this is proven by a calculation
using different sets of the $\eta NN^*$- and $\pi NN^*$-couplings, the
results of which are also shown in Fig~\ref{fig3}. In order to illustrate
the size of the variation of the interaction strength by varying the model
parameters, we quote also the corresponding values of the $\eta N$-scattering
length $a_{\eta N}$ associated with the choice of the coupling constants.
This parameter characterizes the low-energy $\eta N$-interaction and is
quoted in most analyses of the $\eta N$-interaction
(for a review of the $a_{\eta N}$-values see, e.g.~\cite{Green}).
We conclude from Fig~\ref{fig3} that the results are not very sensitive
to the choice of $a_{\eta N}$, so that even with a rather large
value of $a_{\eta N}=(0.91+i0.25)\,$fm (solid curve ``2'' in Fig.\,\ref{fig3}),
we cannot describe the data sufficiently well.


\section{Conclusion}

We have presented theoretical results for the cross section of the reaction
$\gamma d\to \eta np$ in the energy region up to 30 MeV above threshold. The
calculation is performed within a three-body formalism with s-wave separable
$\eta N$- and $NN$-potentials. We found that the first-order rescattering
approximation is insufficient, since the full three-body treatment,
including the full hadronic final state interaction, enhances
significantly at lower energies the $\eta$-meson spectrum while at the higher
energies a less pronounced but still sizeable decrease appears. These
results demonstrate clearly that any "minimal" realistic model for the
$\eta$-production near threshold must be based on a three-body description
of the $\eta NN$-dynamics.

Our calculation explains at least qualitatively the
anomalous behaviour of the experimental $\eta$-meson
yield observed just above threshold.
On the other hand, a quantitative agreement with the experimental results
is not achieved, even with a rather large $\eta N$-scattering length $a_{\eta
N}$=(0.91+i0.25) lying on the boundary of values
allowed by modern analyses. Whether the discrepancy
could be explained by the contribution from the coherent channel
$\gamma d\to \eta d$ is at present an open question.

\begin{figure}
\centerline{\psfig{figure=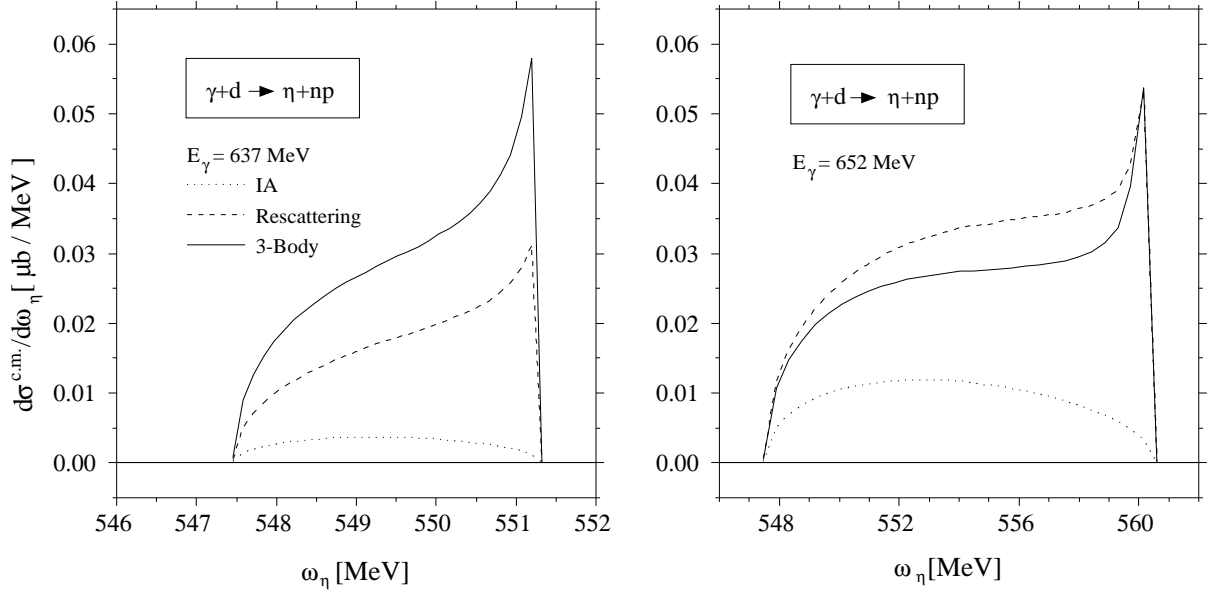,width=16cm,angle=0}}
\vspace{.5cm}
\caption{
The $\eta$-meson spectrum for the reaction $\gamma d\to\eta np$ versus
the total $\eta$ c.m.\ energy $\omega_\eta$ for two lab photon energies.
Notation of the curves: dotted: impulse approximation (IA), dashed:
inclusion of first-order $\eta N$- and $NN$-rescattering, solid:
full three-body calculation.}
\label{fig2}
\end{figure}

\begin{figure}
\centerline{\psfig{figure=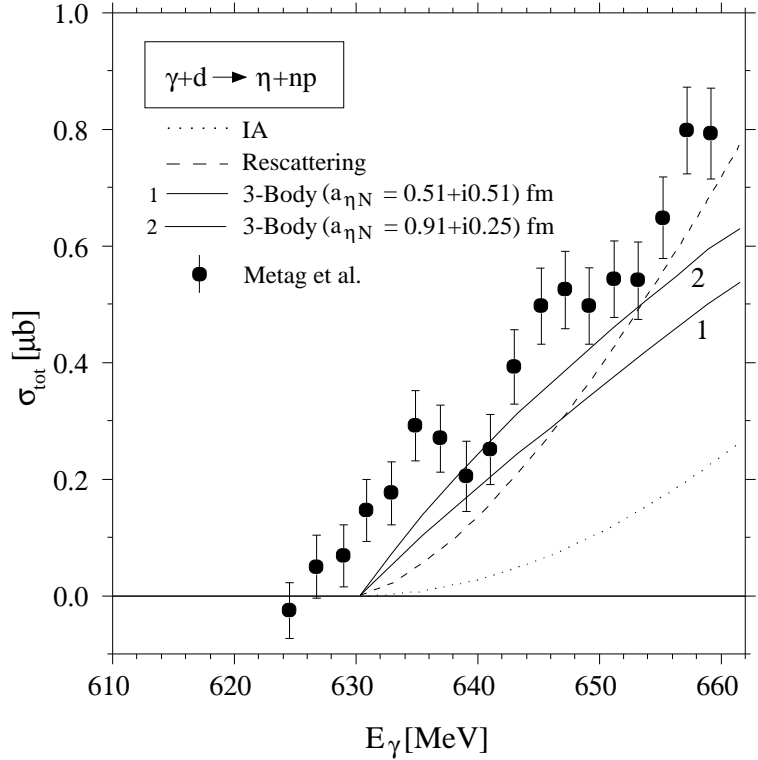,width=10cm,angle=0}}
\vspace{.5cm}
\caption{
Total cross section for the reaction $\gamma d\to\eta np$.
Notation of the curves as in Fig.~\protect\ref{fig2}.
The results obtained within the three-body approach with different
sets of $\eta NN^*$ and $\pi NN^*$ couplings are presented as the
solid curves ``1'' and ``2''.
The corresponding values of the $\eta N$-scattering length
$a_{\eta N}$ are given in the legend.
The inclusive $\gamma d\to\eta X$ data are taken from \protect\cite{Metag}.
}
\label{fig3}
\end{figure}

In view of our results on the $\eta$-meson spectrum as presented in
Fig.~\ref{fig2}, we would like to emphasize the fact that
very useful and interesting information on the mechanism of the low-energy
$\eta$-photoproduction may be obtained from a measurement of
this spectrum. The characteristic feature
of the meson spectrum with its sharp, pronounced maximum close to the high
energy kinematic limit should make
it possible to identify the final state interaction as a major
source of an anomalously large cross section of the reaction
$\gamma d\to\eta np$ close to the threshold. On the contrary,
if there is no simple explanation for the noted shift between the data
and the model predictions, then one has to invoke some
additional unexplored mechanism in $\eta$-photoproduction.





\begin{thebibliography}{99}

\bibitem{Calen}
H.\ C{\'a}len {\it et al}., Phys.\ Rev.\ Lett.\ {\bf 80} (1998) 2069

\bibitem{Metag}
V.\ Metag, Talk at the Baryon-98 conference, Bonn, September 1998

\bibitem{FiAr1}
A.\ Fix, H.\ Arenh{\"o}vel, nucl-th/0006074, Eur. Phys. J. {\bf A} (in print)

\bibitem{Hoff}
P.\ Hoffmann-Rothe {\it et al.}, Phys.\ Rev.\ Lett.\ {\bf 78} (1997) 4697

\bibitem{Krusche2}
B.\ Krusche {\it et al.}, Phys.\ Lett.\ {\bf B358} (1995) 40

\bibitem{Ueda}
T.\ Ueda, Phys.\ Rev.\ Lett.\ {\bf 66} (1991) 297

\bibitem{FiAr}
A.\ Fix, H.\ Arenh{\"o}vel, Z.\ Phys.\ {\bf A359} (1997) 427

\bibitem{Laget}
J.-M.\ Laget, Phys.\ Rep.\ {\bf 69} (1981) 1

\bibitem{AGS}
E.O.\ Alt, P.\ Grassberger, W.\ Sandhas, Nucl.\ Phys.\ {\bf B2} (1967) 167

\bibitem{AAY}
R.\ Aaron, R.D.\ Amado, Phys.\ Rev.\ {\bf 150} (1966) 857

\bibitem{Yamag}
Y.\ Yamaguchi, Phys.\ Rev.\ {\bf 95} (1954) 1628

\bibitem{PDG98}
Particle Data Group, Eur. Phys. J. {\bf C3} (1998) 1

\bibitem{Krusche1}
B.\ Krusche {\it et al.}, Phys.\ Rev.\ Lett.\ {\bf 74} (1995) 3736

\bibitem{Breit}
E.\ Breitmoser, H.\ Arenh\"ovel, Nucl.\ Phys.\ {\bf A612} (1997) 321

\bibitem{Green}
A.M.\ Green, S.\ Wycech, Phys.\ Rev.\ {\bf C55} (1997) 2167

\end{thebibliography}
\end{document}